\documentclass[12pt,preprint]{aastex}

\newcommand{\mdot}{M$_{\odot}$ yr$^{-1}$}

\newcommand{\um}{$\mu$m~}
\newcommand{\ums}{$\mu$m}

\def\kmsMpc{\ifmmode {\rm\,km\,s^{-1}\,Mpc^{-1}}\else
    ${\rm\,km\,s^{-1}\,Mpc^{-1}}$\fi}

\shorttitle{Spitzer Spectra for 10 mJy Sample}
\shortauthors{Houck et al.}

\begin{document}

\title{Spitzer Spectra of a 10 mJy Galaxy Sample and the Star Formation Rate in the Local Universe}

\author{J. R. Houck\altaffilmark{1}, D. W. Weedman\altaffilmark{1},  E. Le
  Floc'h\altaffilmark{2}, and Lei Hao\altaffilmark{1}}

\altaffiltext{1}{Astronomy Department, Cornell University, Ithaca, NY 14853; jrh13@cornell.edu}
\altaffiltext{2}{Steward Observatory, University of Arizona, Tucson, AZ, 85721, and $Spitzer$ Fellow, Institute for Astronomy, University of Hawaii, Honolulu, HI, 96822}

\begin{abstract}
  
A complete flux-limited sample of 50 galaxies is presented having f$_{\nu}$(24\,\um) $>$ 10\,mJy, chosen from a survey with the Multiband Imaging Photometer on $Spitzer$ (MIPS) of 8.2 deg$^{2}$ within the NOAO Deep Wide-Field Survey region in Bootes (NDWFS).  Spectra obtained with the low-resolution modules of the Infrared Spectrograph on $Spitzer$ (IRS) are described for 36 galaxies within this sample; 25 show strong PAH emission features characteristic of starbursts, and 11 show silicate absorption or emission, emission lines, or featureless spectra characteristic of AGN.  Infrared or optical spectral classifications are available for 48 of the entire sample of 50; 33 galaxies are classified as starbursts and 15 as AGN. (There are an additional 19 Galactic stars with f$_{\nu}$(24\,\um) $>$ 10\,mJy in the survey area.)  Using a relation between 7.7\,\um PAH luminosity and star formation rate derived from previous IRS observations of starbursts, the star formation rate per unit volume of the local universe (SFRD) is determined from the complete sample and is found to be 0.008 \mdot Mpc$^{-3}$.  This provides an extinction-free measurement of SFRD independent of optical properties and provides a parameter that can be used for direct comparison to high-redshift starbursts being discovered with $Spitzer$.  Individual sources in the sample have star formation rates from 0.14 to 160 \mdot. The derived value for the local SFRD is about half that of the local SFRD deduced from bolometric luminosities of the IRAS 60\,\um Bright Galaxy Sample, with the deficiency being at lower luminosities and arising primarily from the small number of low luminosity sources in the 10 mJy sample.  The agreement for higher luminosities confirms the validity of using the 7.7\,\um PAH feature as a measure of SFRD in the high redshift universe, where this is often the only indicator available for faint sources.
 
\end{abstract}



\section{Introduction}

Defining flux-limited, complete samples is the basic technique for initial studies of objects discovered by new surveys at various wavelengths.  For example, the  "Bright Galaxy Sample" defined by the Infrared Astronomical Satellite (IRAS) led to fundamental discoveries regarding the natures of luminous infrared galaxies \citep[e.g. ][]{soi89,san03}.  With the capabilities of the Spitzer Space Telescope ($Spitzer$), it becomes possible to define flux limited samples at 24\,\um and obtain infrared spectra for sources which reach factors of $\sim$ 100 fainter fluxes than complete samples defined by IRAS. 

It is important to explore the characteristics of the extragalactic infrared population at these fainter fluxes to determine unbiased luminosity functions and evolution characteristics for various categories of infrared-luminous galaxies.  Such samples are needed for comparison to the large numbers of objects being observed by $Spitzer$ but selected because of previously known optical or IRAS criteria.  It is also especially important to have spectroscopy for brighter, well-defined $Spitzer$ extragalactic samples for comparison to the spectroscopic results for samples with f$_{\nu}$(24\,\um) $\sim$ 1 mJy or fainter, for which hundreds of hours of $Spitzer$ observing time are being invested \citep[e.g. ][]{hou05,wee06a,yan07}. These observations revealed a population of previously unknown, high redshift sources characterised primarily by strong silicate absorption, with a median redshift of 2.1, and we need to understand how these relate to more nearby sources. 

It is also crucial to have statistical distributions of spectral characteristics for $Spitzer$ 24\,\um samples, because modeling of source counts that extends over a wide range of redshift requires knowledge of the spectral shape for objects.  For example, counts have been modeled to deduce the evolution of starbursts by assuming that the source counts arise only from starbursts \citep[e.g. ][]{lag04,cha04}, but it is essential to understand how real sources divide between starbursts and AGN in order to distinguish the cosmic star formation history from the evolution of AGN. 

To define a sample of "Bright Galaxies" discovered by $Spitzer$, we have utilized our survey of 8.2 deg$^{2}$ within the Bootes field of the NOAO Deep Wide-Field Survey
(NDWFS) \citep{jan99}; this $Spitzer$ survey used the Multiband Imaging Photometer for
$Spitzer$ (MIPS) \citep{rie04} to reach 0.3 mJy at 24\,\um and 25 mJy at 70\,\um. The MIPS data were
obtained with an effective integration time at 24\,\um of $\sim$90\,s per sky
pixel, reaching a 5 $\sigma$ detection limit of $\sim$ 0.3\,mJy for unresolved sources, and 40s per sky pixel for the 70\,\um survey, reaching a 5 $\sigma$ detection limit of $\sim$ 25 mJy for unresolved sources.  

We previously reported spectroscopic observations with The Infrared
Spectrograph on $Spitzer$ (IRS)\citep{hou04} for Bootes sources with f$_{\nu}$(24\,\um) $\sim$ 1 mJy \citep{hou05,wee06b}. In the present paper, we use the Bootes survey to define a complete sample of brighter sources, f$_{\nu}$(24\,\um) $>$ 10 mJy, and present new IRS observations for most of this sample. 

Several previous observing programs with the IRS have discovered a substantial population of optically faint sources at z $\sim$ 2 which have PAH emission features in the infrared spectra \citep{lut05,wee06a,yan07,men07}. These features are the signatures of starbursts.  The 7.7\,\um PAH feature is the dominant mid-infrared spectral feature in starbursts at z $\sim$ 2 and is sufficiently strong that it can be measured in faint, high redshift sources with no other quantitative measure of the starburst. The luminosity of this feature provides the potential to measure the star formation rate (SFR) for faint, optically obscured objects at high redshift for which no other indicators of the SFR are available, such as the bolometric luminosity, infrared continuum at rest-frame wavelengths $>$ 10\,\um, H$\alpha$ emission, or rest-frame ultraviolet continuum \citep[e.g. ][]{ken98,mad98,cal07}. %

It is vital to measure the total star formation rate per unit volume of the universe - the star formation rate density (SFRD) - at high redshifts using the various observations enabled by $Spitzer$ to obtain a result which is independent of other methods used to track the cosmic SFRD as a function of epoch in the universe \citep{lef05,tak05,man07}.  Because the PAH luminosity is much less subject to extinction than rest-frame optical or ultraviolet and is the best mid-infrared indicator observable at high redshift, it may prove to be the primary measure of starburst luminosity in faint, distant sources discovered within $Spitzer$ surveys.  


The first step in enabling this determination is a calibration of SFR as it relates to PAH luminosity, and we present an empirical calibration derived from the global characteristics of classical starburst galaxies observed with the IRS.  In order to determine if this calibration is applicable to the SFRD  averaged over all starbursts, we determine the local SFRD using our new Bootes 10 mJy sample and compare it to other estimates of the local SFRD. 

Previous observations of the Bootes survey region have produced large amounts of multiwavelength data, including deep optical measurements in $B_W$, $R$, and $I$ \citep{jan99}, a survey \citep{eis04} with all four bands of the $Spitzer$ Infrared Array Camera \citep{faz04}, a 5 Ks survey with the Chandra X-ray Observatory \citep{brn06}, and optical spectroscopy of many sources in the AGN and Galaxy Evolution Survey \citep{coo06}.  All of these data potentially enable a detailed comparison of various multiwavelength properties of all sources within the 10 mJy sample.  We do not undertake such an analysis in the present paper because our objective is to present only the sample and the new IRS results.   The extensive multiwavelength data are not utilized for our determination of the local SFRD using the infrared spectra, because we want to determine the validity of this technique for sources which have no data other than infrared spectra.

\section{Sample Definition, New IRS Observations, and Data Analysis}

The Bootes bright source sample is defined only by f$_{\nu}$(24\,\um) $>$ 10 mJy.  For resolved sources with several components in the same galaxy, this criterion is for the total flux of the galaxy. There are 69 sources within the 8.2 deg$^{2}$ Bo\"{o}tes field which satisfy this flux criterion.  Of these 69, 19 are optically bright Galactic stars, as classified on the Digitized Sky Survey and the Digitized First Byurakan Survey \citep{mic07}.  The remaining 50 extragalactic sources are listed in Table 1, which summarizes the characteristics for the complete sample, including sources within our program and sources in other $Spitzer$ programs.  

Of the 50 sources, we describe new IRS results for 26 sources within our program and 10 from the $Spitzer$ public archive (program 20113, H. Dole, in preparation).   Adding optical spectral classifications from the Sloan Digital Sky Survey (SDSS) \citep{gun98,yor00} allows the spectroscopic classification of the sources as starbursts or AGN for 48 of the 50 sources in the complete sample.  $Spitzer$ spectroscopic observations were made with the IRS\footnote{The IRS was a collaborative venture between Cornell
University and Ball Aerospace Corporation funded by NASA through the
Jet Propulsion Laboratory and the Ames Research Center.} Short Low module in
orders 1 and 2 (SL1 and SL2) and with the Long Low module in orders 1 and 2 (LL1 and
LL2), described in \citet{hou04}.  These give low resolution spectral
coverage from $\sim$8\,\um to $\sim$35\,\um.  

For our new observations, sources were placed on
the slit by using the IRS peakup mode with the blue camera.  All images when the source was in one of the two nod positions on each
slit were coadded to obtain the source spectrum.  The background which was subtracted was determined from coadded background images that added both nod positions having the source in the other slit (i.e., both nods on the LL1 slit when the source is in the LL2 slit produce LL1 spectra of background only).  
The difference between coadded source images minus coadded background images was used for the spectral extraction, giving two independent extractions of the spectrum for each order.  These independent extractions were compared to reject any highly outlying pixels in either
spectrum, and a final mean spectrum was produced.  

Extraction of
source spectra was done with the SMART analysis package \citep{hig04}, beginning with the bcd products of version 13.0 of the $Spitzer$ flux calibration pipeline.  Final spectra were boxcar-smoothed to the approximate resolution of the different IRS modules (0.2\,\um for SL1 and SL2, 0.3\,\um for LL2, and 0.4\,\um for LL1).  Spectra from our new observations of starburst sources in Table 1 are illustrated in Figure 1 (illustrated spectra are truncated at 25\,\um because of the absence of PAH features beyond that wavelength).

\section{Discussion}

\subsection{Characteristics of the Bootes 10 mJy Sample}

Because this sample is defined by infrared flux density, it provides an overall distribution of the types of extragalactic sources which are encountered in an unbiased examination of the $Spitzer$ infrared sky.  The sources discovered illustrate examples of all representative infrared spectra from extragalactic sources, including strong PAH emission, strong silicate absorption, strong emission lines, and featureless power laws. 

We utilize a simple classification for IRS spectra, which divides sources into those with infrared spectra characterized by strong PAH emission features  or spectra with silicate absorption, strong emission lines, or featureless continua.  The PAH spectra are assigned to starbursts \citep[e.g. ][]{bra06} and the remaining spectra to AGN \citep[e.g. ][]{wee05}. Sources can be composite, but we do not attempt deconvolution into AGN and starburst components; we classify only by whether PAH features are present, or not.  Quantitative classifications according to the scheme of \citet{spo07} are also assigned.   The classification of the IRS spectrum is given in Table 1 for sources with available IRS spectra.  Sources without IRS spectra but with SDSS spectra are also assigned as starburst or AGN based on the SDSS spectra. 
  
Of the 48 sources with spectroscopic classifications, infrared or optical, 33 are starbursts and 15 are AGN.  Several sources with an optical classification as Seyfert 2 show strong PAH emission in the infrared; in such cases, we include these as starbursts.  There are 25 sources in Table 1 for which IRS spectra show PAH emission, and we will discuss these in more detail. The remaining 11 sources having IRS spectra display silicate absorption, possible silicate emission, emission lines, or featureless spectra and so are attributed to AGN.  Because of the variety of these AGN spectra and the small number of sources within different AGN categories, we defer presentation and detailed analysis of the AGN spectra until we have a larger sample.  For the remainder of this paper, we discuss only the starburst sources.  There is a sufficient number of these to derive meaningful conclusions regarding star formation at low redshifts, which we present in section 3.3. 

\subsection{Starburst Galaxies in the 10 mJy Sample}

The subset of 15 sources from the Bootes 10 mJy sample for which we have obtained new IRS spectra that show PAH features (indicating starbursts) is listed in Table 2 (the archive observations from program 20113 are not included in Table 2).  This Table gives all of the spectroscopic characteristics which were measured from our spectra, except for redshifts which are given in Table 3.  IRS redshifts are determined from PAH emission features, assuming rest wavelengths of 6.2$\mu$m, 7.7$\mu$m, 8.6$\mu$m, and 11.3$\mu$m.  The IRS redshifts typically agree to within 0.0012 of the optical redshift from SDSS. 
All spectra in Table 2 are shown in Figure 1 (truncated at 25\,\um because of the lack of PAH features beyond that wavelength).  

Most sources show PAH features with large equivalent width (EW).  The most common classifier of PAH strength is the EW of the 6.2\,\um PAH feature, which is listed in Table 3 for all objects in the sample which show this feature.  The weakest feature detected has EW of 0.05$\mu$m. Classification of sources according to the EW criteria of \citet{spo07} are in Table 1.

We desire a quantitative classification to place these objects in context of other starbursts, especially the classical starburst galaxies whose IRS spectra are described by \citet{bra06} and the Blue Compact Dwarfs (BCDs) described by \citet{wu06}.  For this purpose, the diagnostics we use are the ratios among [SIV] 10.5\,\um, PAH 11.3\,\um, and [NeII] 12.8\,\um.  These ratios are the best available to compare with the previous starburst and BCD spectra, because these features all arise within the SL orders.  Although the galaxies in the 10 mJy sample are unresolved and so are completely included within the IRS slits, the starburst galaxies and BCDs in the comparison studies are resolved, so that significant but uncertain aperture corrections arise when comparing features between SL and LL modules, which have different slit widths.  

These ratios need to be measured on low resolution spectra which are the same as we have for the Bootes objects, but low-resolution measures of these features were not previously published. We have, therefore, re-extracted all of the SL1 spectra for the 10 BCDs in Table 4 of \citet{wu06} and for the 22 starburst galaxies in \citet{bra06}, using the same extraction process
as for the Bootes 10 mJy sources.  The comparison of these 3 samples is illustrated in Figure 2.  

This Figure illustrates that starbursts and BCDs are separated by the [SIV]/[NeII] ratio.  It also illustrates that all of the Bootes PAH sources are in the regime of classical starbursts, except for two objects which have [SIV]/[NeII] ratios previously observed only in BCDs.  (One Bootes source did not have the necessary SL observations for classification.)  The two sources within the BCD regime are sources 25 and 30 in Table 2.  Source 25 has a Seyfert 2 component, according to the SDSS optical classification, which explains the strong [SIV], so this source is not a BCD.  Source 30 is confirmed as a true BCD in the SDSS spectra, and this source also has the lowest luminosity of all PAH sources.  It is also classified as a BCD by \citet{ros06} because of the strong emission lines in the objective prism survey of the Kitt Peak International Spectroscopic Survey (KISS).

\subsection {PAH Luminosity and Star Formation Rate}

Knowledge of the total star formation rate per unit volume of the universe, or star formation rate density (SFRD), as a function of age of the universe is crucial for understanding the formation and evolution of galaxies.  Various studies have attempted to measure cosmic evolution of the SFRD by locating starburst galaxies using ultraviolet, optical, and near-infrared techniques \citep[e.g. ][]{mad98,lef05,tak05,man07}. This is done by scaling the ultraviolet luminosity to a star formation rate using assumed initial mass functions.  The primary source of uncertainty is the extinction uncertainty in the rest-frame optical or ultraviolet, and this is a major uncertainty for the majority of luminous starbursts which are heavily obscured.  To avoid this uncertainty, and to utilize a measure of the bolometric luminosity of the starburst, it is important to determine the SFRD from infrared luminosities of galaxies \citep [e.g. ][]{bel05, lef05, cha05, cap07}.  

While these previous efforts use measures of the infrared continuum emission from dust, a potential alternative technique is to use the flux of PAH emission features characteristic of infrared spectra from starbursts as a measure of the star formation rate within a galaxy (SFR).  
Approximate indicators of the validity of this approach were found using the Infrared Space Observatory broad-band photometry in the LW2 band which centers on the 7.7\,\um PAH feature for low redshift galaxies \citep{for04}.  A broad-band measure can be an accurate determination of the 7.7\,\um flux for sources with redshifts such that this feature is within the band observed.  However, for sources covering a range of redshifts, the transformation from broad-band flux to 7.7\,\um flux depends on assuming a template for the spectrum that relates continuum and PAH fluxes \citep{cap07}. Having spectra that allow a direct measure of the 7.7\,\um feature regardless of redshift avoids such assumptions and is the approach that we utilize. 

While the precise natures of the PAH features depend on the physical conditions within the star-forming region \citep{dra07}, and the PAH spectrum varies among star-forming regions within the same galaxy \citep{smi07}, there is an unquestionable empirical association between the presence of an optically-classified starburst and the presence of PAH features in the infrared spectrum \citep[e.g. ][]{gen98,rig00,bra06}.  Furthermore, the integrated PAH spectra of starburst galaxies are notably uniform \citep{bra06}. It is reasonable, therefore, to exploit the empirical correlation between PAH features and starbursts to determine an empirical relation between PAH luminosity and SFR.  

Measuring the total fluxes in PAH features is challenging, even in spectra with high signal to noise ratios (S/N), such as in Figure 1.  The features are sometimes broad, and the definition of the appropriate underlying continuum is challenging and complex, especially if silicate absorption is also present \citep{spo07}. The strongest PAH emission feature is at rest frame 7.7\,\um, but the underlying continuum that should be subtracted from the emission feature is especially difficult to define if there is no observed continuum baseline at longer wavelengths.  This occurs for faint sources with z $\ga$ 2 for which the IRS spectra do not include rest frame wavelengths $\ga$ 11\,\um.   

We desire to relate our measurements to faint sources at high redshifts, so we wish to utilize the parameter most easily measured for such sources.  Because of poor S/N in these faint sources and the difficulty in defining the continuum, the parameter we utilize for PAH luminosity is $\nu$L$_{\nu}$ (7.7$\mu$m), where 
L$_{\nu}$ (7.7$\mu$m) is determined from the flux density at the peak of the 7.7$\mu$m feature.  If significant underlying continuum is present, an estimate of that continuum can be subtracted to obtain the PAH luminosity.  In the spectra in Figure 1 and the starbursts in \citet{bra06}, the underlying continuum is not significant compared to the peak of this feature; the continuum which is present is also probably associated with the same starburst that produces the PAH emission.  For the present paper, therefore, we utilize only the peak f$_{\nu}$ (7.7$\mu$m) with no correction for the underlying continuum.  

Because the SFR has been calibrated as a function of total bolometric luminosity for sources powered by starbursts \citep{ken98}, we desire to relate the $\nu$L$_{\nu}$ (7.7$\mu$m) to the bolometric luminosity, and thereby to the SFR.  
We can determine an empirical calibration by measuring $\nu$L$_{\nu}$ (7.7$\mu$m) for the 22 starbursts in \citet{bra06}, and comparing to the $L_{ir}$ tabulated there as determined from the IRAS fluxes \citep{san03}. Brandl et al. include aperture corrections for the finite size of the IRS slits compared to the IRAS total fluxes, and we apply these corrections.  The resulting comparison is shown in Figure 3.  The best linear fit which transforms $\nu$L$_{\nu}$ (7.7$\mu$m) to $L_{ir}$ without any luminosity dependence is shown in Figure 3, log $L_{ir}$ = log[$\nu$L$_{\nu}$ (7.7$\mu$m)] + 0.78.  The one sigma dispersion in this relation is 0.2, indicating that $\nu$L$_{\nu}$ (7.7$\mu$m) can predict $L_{ir}$ to within a factor of 1.6 for a starburst over a range of $\sim$ 100 in luminosity.  Using the relation from \citet{ken98}, this yields that log[SFR] = log[$\nu$L$_{\nu}$ (7.7$\mu$m)] - 42.57, for SFR in units of \mdot.  

This transformation from a PAH luminosity to a SFR depends on relating the SFR to the bolometric luminosity, assuming that the bolometric luminosity is given by $L_{ir}$, and calibrating the PAH luminosity to $L_{ir}$.  A completely independent estimate of SFR derives from the luminosity of the [Ne II] and [Ne III] lines \citep{ho07}.  These luminosities have been related to the luminosities of the 6.2$\mu$m plus 11.3$\mu$m PAH features by \citet{far07}, who then determine a SFR measure from the total luminosities of these PAH features.  Their measure can be transformed to $\nu$L$_{\nu}$ (7.7$\mu$m) by empirically relating $\nu$L$_{\nu}$ (7.7$\mu$m) to L(6.2$\mu$m + 11.3$\mu$m).  From the starbursts in \citep{bra06}, $\nu$L$_{\nu}$ (7.7$\mu$m)/L(6.2$\mu$m + 11.3$\mu$m) = 40, with a dispersion of only 10\%.  With this transformation, the SFR measure in \citet{far07} becomes log[SFR] = log[$\nu$L$_{\nu}$ (7.7$\mu$m)] - 42.52.  The agreement to within 10\% of this independent calibration based on Ne emission lines to the calibration derived above based on $L_{ir}$ is certainly fortuitous, but it does provide a confirmation of the relation that we have adopted. 

As a further check to estimate the overall global validity of the calibration we have adopted between $\nu$L$_{\nu}$ (7.7$\mu$m) and SFR, we use the PAH luminosity in the starbursts of the complete 10 mJy sample to determine the integrated SFRD of the local universe.  This result is then compared to independent determinations of the local SFRD to determine the systematic uncertainty in this use of PAH luminosity to determine star formation rates.

\subsection {Space Densities and Star Formation Rate Density in the Local Universe}

The luminosities $\nu$L$_{\nu}$(7.7$\mu$m) of the starbursts in Table 2 are given in Table 3.  These luminosities range from 5.2 x 10$^{41}$ ergs s$^{-1}$ to 2.4 x 10$^{45}$ ergs s$^{-1}$.  For comparison, $\nu$L$_{\nu}$(7.7$\mu$m) of the prototype starburst NGC 7714 is 4.4 x 10$^{43}$ ergs s$^{-1}$\citep{bra06}, and $\nu$L$_{\nu}$(7.7$\mu$m) for the most luminous starbursts discovered by $Spitzer$ at z $\sim$ 2 is $\sim$ 10$^{46}$ ergs s$^{-1}$, \citep[e.g. ][]{yan05,wee06a}.

Most of the starburst sources are detected at 70$\mu$m, so the ratio f$_{\nu}$(24$\mu$m)/f$_{\nu}$(70$\mu$m) can be measured and is shown in Figure 4 compared to the PAH luminosities. The median value for the ratio is $\sim$ 0.09, with no dependence on luminosity.  If the slope of the continuum measured by this ratio is a measure of dust temperature, or mixtures of dust temperatures, these results imply that such temperatures do not relate to the luminosity of the starburst. This is an indication that the uniformity of the PAH strength relative to the underlying continuum arises from a simple scaling of multiple, similar starbursts within a source. 

The SFRs derived from $\nu$L$_{\nu}$(7.7$\mu$m) are also in Table 3 and range from 0.14 to 160 \mdot.  These SFRs are generally much larger than those which would be derived from the relation between L(H$\alpha$) and SFR \citep{ken94} if no corrections are applied for extinction at H$\alpha$.  Comparison of SFR derived from $\nu$L$_{\nu}$(7.7$\mu$m) and from L(H$\alpha$) are shown in Table 3 for those sources having f(H$\alpha$) in \citet{jan05} from the KISS survey.  The H$\alpha$-derived SFRs are always smaller, often by a factor of $\sim$ 10, except for one object - the low-luminosity BCD, number 30.  For all other starburst galaxies, the results imply that the majority of the radiation from the starburst has been absorbed by dust. 

Space densities in Table 3 are derived from the standard V$_{max}$ technique, determining the maximum volume which a source could inhabit and remain within the 10 mJy sample. The SFRDs derive from the space densities. 

Using the sample of 25 galaxies in Table 3, we can derive the total local space densities and resulting SFRDs as a function of luminosity.  Results are in Table 4. The total local SFRD determined only from the sources in Table 3 is 0.0063 \mdot Mpc$^{-3}$, and this value decreases to 0.0057 if the 4 starbursts in Table 3 at z $>$ 0.2 are excluded.  This value is strictly a sum of SFRDs for all sources observed.  Individual luminosity bins are subject to large statistical uncertainties because of the small numbers of sources in each bin.  Numbers of sources in each bin are listed in Table 4, and the statistical uncertainties are illustrated in Figure 5.

Some correction for incompleteness is necessary, because the galaxies in Table 3 with observed $\nu$L$_{\nu}$(7.7$\mu$m) are only 25 of the 33 starburst sources in the complete sample, adopting the optical classifications for starbursts in Table 1 for sources without infrared spectra.  If the remaining starbursts are similar to those in Table 3, this incompleteness correction would raise the estimate of the local SFR (z $<$ 0.2) to $\sim$ 0.008 \mdot Mpc$^{-3}$.  

It is notable that 4 sources classified as starbursts from the presence of PAH features are classified as Sy 2 galaxies from the optical SDSS spectra.  The infrared spectra are clearly composite because the PAH features in these objects have weaker equivalent widths than in the starbursts.  Source 28 is also probably such a composite although it does not have an optical spectrum for classification; the PAH feature has low equivalent width, and the source has very high luminosity. A small correction could be made to the SFR because of an overestimate of the SFR arising by measuring only the peak $\nu$L$_{\nu}$(7.7$\mu$m) within these starbursts in Table 3 that are also Sy 2 galaxies; some of the underlying continuum arises from the AGN, although all of the continuum is included in our calibration for SFR compared to $\nu$L$_{\nu}$(7.7$\mu$m).  We do not attempt to apply a correction for the presence of such composite AGN/starburst spectra, however, because the uncertainty arising from this is small compared to the statistical uncertainties arising from the small sample size. 

This result for the total local SFRD of  0.008 \mdot Mpc$^{-3}$ is smaller than the result previously derived using the luminosity function derived at 60\,\um from the IRAS Bright Galaxy Sample (BGS).  For example, the compilation of \citet{lef05}, which adopts the 60\,\um luminosity function of \citet{tak05}, yields an infrared-derived local SFRD of 0.013 \mdot Mpc$^{-3}$.  

To determine the total local SFRD, the infrared-derived value, which represents obscured star formation, should be added to the value derived from the non-extincted ultraviolet luminosity from starbursts, which represents the primary radiation that escapes without dust absorption.  Ultraviolet observations with the Galaxy Evolution Explorer (GALEX) yield a observed local SFRD, not corrected for extinction, of 0.003 M$_{\odot}$ yr$^{-1}$Mpc$^{-3}$ \citep{igl06}, which indicates a total local SFRD of 0.016 \mdot Mpc$^{-3}$ when both obscured and unobscured star formation are accounted for.  A similar factor for infrared-derived SFRD compared to ultraviolet-derived SFRD is found when individual galaxy templates are used with models of evolving star formation \citep{bel05}. 

These empirical results indicate, therefore, that our use of the 7.7$\mu$m flux to determine the local SFRD from our 10 mJy sample of starburst galaxies  yields a value which is $\sim$ 50\% of the value derived from more sophisticated and extensive analyses that sum the infrared and ultraviolet luminosities of much larger galaxy samples.  To understand more about the sources of this difference, we make a more detailed comparison to the SFRDs derived from the IRAS BGS.

\subsection{Comparison with IRAS Bright Galaxy Sample}

Derivations of the SFR for dust-obscured starbursts depend on transformation of the 60\,\um luminosities of IRAS sources to bolometric luminosities, $L_{ir}$, which are transformed in different ways by different authors.  The calibration between IRAS fluxes and $L_{ir}$ which we adapt from \citet{bra06} for the 22 starburst galaxies used for comparing $\nu$L$_{\nu}$(7.7$\mu$m) to $L_{ir}$ is the same as the calibration used in \citet{san03}.  Any differences in results for space densities and SFRs between our 10 mJy sample compared to the BGS in Sanders et al. derive, therefore, from systematic errors in our determination of $\nu$L$_{\nu}$(7.7$\mu$m) to $L_{ir}$ or from deficiencies in the 10 mJy sample compared to the larger BGS sample. 

We reproduce in Table 4 the space densities and SFRs from \citet{san03} for the BGS for comparison to our 10 mJy results. ($L_{ir}$ in Table 4 are in L$_{\odot}$.)  This comparison is also shown in Figure 5, which includes upper limits and statistical uncertainties for the 10 mJy sample of starbursts.  The total SFRD from the summed BGS space densities and SFRDs in Table 4 would be 0.018 \mdot Mpc$^{-3}$.  While larger by $\sim$ 50\% than the value which others derive using the same 60\,\um luminosity function \citep[e.g. ][]{lef05,xu06}, this difference arises from different methods of converting 60\,\um luminosities, $L_{60}$, to $L_{ir}$.  Because our calibration for SFR began with the Sanders et al. values for $L_{ir}$, any differences between the samples in Table 4 do not arise from different assumptions about relating $L_{ir}$ to $L_{60}$.  The conversion of luminosity to SFR given in Table 4 for the BGS and the 10 mJy sample also derives from the same calibration of SFR to $L_{ir}$ \citep{ken98}.  To understand differences in the local SFRDs, therefore, we need only to understand why the IRAS BGS yields space densities for starburst galaxies that are different than the space densities we derive from the $Spitzer$ 10 mJy sample.

The differences in Table 4 and Figure 5 between the space densities and resulting SFRDs from the IRAS sources and from our 10 mJy sources clearly depend on source luminosity.  The 10 mJy sample is deficient at lower luminosities.  This is despite the fact that the median luminosities of the samples are very similar; the IRAS median is log $L_{ir}$ = 10.65, and for the 10 mJy sources in Table 3 is log $L_{ir}$ = 10.9.  The IRAS space densities for all bins with log $L_{ir}$ $<$ 10.5 are greater than those for the 10 mJy sample, whereas the two samples give similar results for the higher luminosity bins with log $L_{ir}$ $>$ 10.5.

Most of the differences appear to arise from the very small number of sources in the 10 mJy sample.  Four of the 6 bins with log $L_{ir}$ $<$ 10.5 have no sources discovered; the upper limits shown in Figure 5 are based on a limit of one source in each bin.  By contrast, the single bin with the most 10 mJy sources (10.5 $<$ log $L_{ir}$ $<$ 11.0 with 10 sources) yields a space density that agrees to within $\sim$ 10\% of the BGS result.  For all bins with log $L_{ir}$ $>$ 10.5, the summed results agree identically, with IRAS giving 31 x 10$^{-5}$ starburst galaxies Mpc$^{-3}$, and the 10 mJy sample also yielding 31 x 10$^{-5}$  Mpc$^{-3}$.

These comparative results indicate that for high luminosities, which are the luminosity bins within which sufficient 10 mJy sources are found for meaningful statistics, the calibration of PAH luminosity to bolometric luminosity gives consistent values for local space densities, and thereby for local SFRDs, as given by previous measures from the BGS.  This is an encouraging result, because our eventual objective in testing this calibration is to compare SFRDs in the local universe to SFRDs at high redshift, where only the high luminosity sources can be observed with $Spitzer$.   

Nevertheless, it is important to consider possible reasons for the discrepancies between BGS and 10 mJy samples at low luminosities.  In order to include most of the 10 mJy starbursts in a determination of the local SFRD, we defined "local" to mean z $<$ 0.2.  The mean redshift of the 25 galaxies in Table 3 is 0.115, which is about a factor of 10 greater than for the BGS.  As a result, we are comparing very different volumes of the universe.  As also 
emphasized by the very small number of sources in both samples in the lowest luminosity bins of the 10 mJy sample, a reasonable explanation for the differences is that accidents of cosmic variance mean that the 10 mJy sample did not include a volume containing many low luminosity starbursts.  This test can be improved by increasing the sample to other areas surveyed by $Spitzer$.   

Another possibility is that a 24\,\um survey to find starbursts and estimate their $L_{ir}$ from the strength of PAH emission leads to a sample of starbursts which, for some reason, preferentially omits low luminosity sources compared to sources discovered by IRAS.  Because our determination of $L_{ir}$ for the 10 mJy sample is done using the calibration with $\nu$L$_{\nu}$(7.7\,\um), perhaps this calibration systematically is different for lower luminosity sources.  This could happen, for example, if low luminosity starbursts have cooler dust and higher ratios of f$_{\nu}$(60\,\um)/f$_{\nu}$(24\,\um), so that lower luminosity starbursts are preferentially revealed in the IRAS 60\,\um surveys compared to the 24\,\um surveys.  There is no evidence of such a trend in Figure 4 comparing f$_{\nu}$(24$\mu$m) with f$_{\nu}$(70$\mu$m), and the empirical calibration for starbursts in Figure 3 covers all of the lower luminosity range in Table 4.  Based on the limited data available in the 10 mJy sample, we have no evidence for any systematic spectral differences among starbursts as a function of luminosity. 

It is known, however, that low luminosity starbursts of low metallicity (BCDs) generally have weak PAH features \citep{wu06} compared to the continuum strength.  For such sources, our adopted calibration between $\nu$L$_{\nu}$(7.7\,\um) and $L_{ir}$ would overestimate $L_{ir}$ because the calibration assumes negligible contribution by the continuum to $\nu$L$_{\nu}$(7.7\,\um).  Without correcting the PAH luminosity for the underlying continuum, BCDs with weak PAH features would be incorrectly assigned to higher luminosity bins of $L_{ir}$.  Only one of the 10 mJy galaxies in our sample is a BCD (number 30) so reassigning this object to a lower luminosity bin would not explain the systematic deficiency of sources with low $L_{ir}$; however, the presence of only one low-luminosity BCD indicates the statistical deficiency of the 10 mJy sample.

\section{Summary and Conclusions}

We present a complete sample of 50 galaxies discovered by $Spitzer$ within 8.2 deg$^{2}$ having f$_{\nu}$(24\,\um) $>$ 10\,mJy.  IRS spectra are available for 36 galaxies, of which 25 show strong PAH emission features characteristic of starbursts, and 11 show spectral characteristics of AGN.  Of the complete sample of 50, 48 galaxies have classifications from either infrared or optical spectra, of which 33 are starbursts and 15 are AGN. This result indicates that unbiased samples of $Spitzer$ sources selected with no criteria other than f$_{\nu}$(24\,\um) are dominated by starbursts. 

The starburst sample is used to derive the star formation rate in the local universe (z $<$ 0.2) based only on the luminosities of the 7.7\,\um PAH feature, using an empirically determined relation between $\nu$L$_{\nu}$(7.7\,\um) and $L_{ir}$ derived from a previously observed sample of starbursts \citep{bra06}.   The result gives a local star formation rate density of 0.008 M$_{\odot}$ yr$^{-1}$ Mpc$^{-3}$, which is about a factor of two lower than previous estimates derived from $L_{ir}$ using the IRAS Bright Galaxy Sample. Comparison of the IRAS and 10 mJy samples indicates that the deficiency in the measured local SFRD arises because our 10 mJy sample finds far fewer galaxies for $L_{ir}$ $<$ 10.5 (units of L$_{\odot}$), probably because of the small size of the 10 mJy sample.   For $L_{ir}$ $>$ 10.5, the derived space densities and resulting SFRDs are similar between the IRAS and 10 mJy samples.  

The primary motivation for using $\nu$L$_{\nu}$(7.7\,\um) to derive SFRs is the need to apply this technique to faint $Spitzer$ sources being discovered at z $\sim$ 2, for which $\nu$L$_{\nu}$(7.7\,\um) is the best measurable parameter that relates to SFR.   Use of this parameter gives a method for comparing SFRDs in the universe at different epochs.  Our confirmation that this technique is valid for determining the global SFRD for high luminosity galaxies in the local universe indicates that $\nu$L$_{\nu}$(7.7\,\um) should also give a reliable measure of SFRDs at high redshifts, where only the most luminous sources are detectable. 

\acknowledgments
We thank D. Devost, G. Sloan and P. Hall for help in improving our IRS spectral analysis, A. Mickaelian for classification of bright stars in the survey, K. Brand for help in organizing the 10 mJy catalog, and B. T. Soifer for discussions.  This work is based primarily on observations made with the
Spitzer Space Telescope, which is operated by the Jet Propulsion
Laboratory, California Institute of Technology under NASA contract
1407. Support for this work by the IRS GTO team at Cornell University was provided by NASA through Contract
Number 1257184 issued by JPL/Caltech.

\clearpage
\pagestyle{empty}

\begin{deluxetable}{lccccccc} 
\tablecolumns{8}
\tabletypesize{\footnotesize}

\tablewidth{0pc}
\tablecaption{The Bootes 10 mJy Sample}
\tablehead{
  \colhead{Number}& \colhead{Source Name\tablenotemark{a}}& \colhead{AOR}\tablenotemark{a}& \colhead{program}& \colhead{time\tablenotemark{b}}& \colhead{f$_{\nu}$(24$\mu$m)\tablenotemark{c}}& \colhead{f$_{\nu}$(70$\mu$m)\tablenotemark{d}}& \colhead{class\tablenotemark{e}}\\ 
  \colhead{}& \colhead{}& \colhead{}& \colhead{}& \colhead{s}& \colhead{mJy}& \colhead{mJy}& \colhead{}
}
\startdata

1 &SST24 J142640.65+321151.7 &\nodata &\nodata &\nodata & 19.5 & 20 & QSO\tablenotemark{f}\\      
2 &SST24 J142646.63+322125.6 & 17539328 & 30121 & 240,480 & 12.1(11.5) & 120 & PAH,1C\\
3 &SST24 J142824.91+322340.7 &\nodata &\nodata &\nodata &43.0(r) &41 &starburst\tablenotemark{f}\\
4 & SST24 J142629.15+322906.7 & 16348416 & 16 &240,480 & 12.7(13.3) & 177 & PAH,1B\\
5 & SST24 J142623.91+324436.0 & 14084352 & 20113 & 120,120 &  12.1 & 255 & PAH,1B\\ 
6 & SST24 J143115.23+324606.2 & 16348672 & 16 &120,240 & 23.4(23.6) & 165 & PAH,1C\\
7 & SST24 J143156.40+325138.1 & 16346880 & 16 &240,480  & 13.4(13.8) & $<$ 15 &Si emission?,1A?\\
8 & SST24 J143024.49+325616.5 & 14082304 & 20113 & 120,60 & 40.0 &380 &starburst\tablenotemark{f}\\
9 & SST24 J143205.63+325835.2 & 16160256  & 15 & 240,480 & 16.6(14.5) & 106 &Si absorption,2A\\
10 & SST24 J143602.53+330753.7 & \nodata &\nodata &\nodata & 53.7 & 473 &starburst\tablenotemark{f}\\ 
11 & SST24 J143445.35+331346.2 & 14083584 & 20113 & 120,240 & 14.2 & 205 & starburst\tablenotemark{f}\\ 
12 & SST24 J143125.46+331349.8 &14080768 & 20113 & 120,60 & 53.5(r) & 605 & PAH,1C\\
13 & SST24 J142847.19+332316.1 &\nodata &\nodata& \nodata &30.0(r) &520 &starburst\tablenotemark{f}\\ 
14 & SST24 J142652.95+332323.1 & 10493696 & 3231 & \nodata & 10.8 & \nodata &QSO\tablenotemark{f} \\
15 & SST24 J142659.15+333305.1 & 14084096 & 20113 & 120,60 & 14.5 & 90? &PAH,1C\\
16 & SST24 J142658.80+333314.0 & 16347648 & 16,20113 & 120,480 & 13.4 & 340? & PAH,1C\\
17 & SST24 J143628.15+333358.1 & 14087168 & 20113 & 240,240& 10.8 & 192 &\nodata\\ 
18 & SST24 J143154.92+333816.2 & \nodata & \nodata &\nodata & 13.0 & 170 &starburst\tablenotemark{f}\\ 
19 & SST24 J143156.25+333833.3 & 14081536 & 20113 & 120,60 & 48(r) & 585 & PAH,1C\\
20 & SST24 J142543.59+334527.6 & 16349696  & 16 &120,240 & 45.7(48.1) & 192 &PAH,1C\\
21 & SST24 J143310.33+334604.5 & 17539584 & 30121 &240,480 & 10.3(10.1) & 30 &featureless\\
22 & SST24 J143409.54+334649.4& 16347136 & 16 &240,480 & 12.9(12.5) & $<$ 15 & emission lines?,1A\\
23 & SST24 J143632.01+335230.7 & 17538560 &30121 &240,480 & 10.5(11.2) & 143 & PAH,1C\\
24 &  SST24 J142552.71+340240.2 & 14087680 & 20113 & 240,240 & 11.3 & 215  & PAH,1B\\
25 & SST24 J143232.52+340625.2 & 17539072 & 30121 &240,480 &10.2(10.1) & 19 &PAH,1B\\
26  & SST24 J143105.67+341232.7& 17538048 & 30121& 240,480 & 11.4(11.4) & 22 &weak Si absorption?,1A\\
27  & SST24 J143132.17+341417.9 & 16347392 & 16  &240,480  & 12.2(11.9) & 35 & featureless,1A\tablenotemark{h}\\
28  & SST24 J143157.96+341650.1& 16348160 & 16 &240,480 &19.6(19.7) & 37 & emission lines,1A\tablenotemark{g}\\
29 & SST24 J142417.44+342046.7 & \nodata &\nodata &\nodata & 34.2 & 227 & starburst\tablenotemark{f}\\ 
30 & SST24 J143120.00+343804.2 & 16349184 & 16 &120,240 & 26.6(27.2) & 153& PAH,1C\\
31 & SST24 J143631.98+343829.3 & 16349952 & 16 &120,240 & 47.5(48.5) & 470 &PAH,1A\\
32 & SST24 J143126.81+344517.9 & 17538816 & 30121 &240,480 & 10.6(10.0) & 130 & PAH,1C\\
33 & SST24 J143728.80+344547.6 & 16348928  & 16 &120,240 & 23.5(24.6) & 82 & Si absorption,2A\\
34 & SST24 J142554.57+344603.2 & 16350208 & 16 &120,240 & 51.8(55.4) & 620&PAH,1C\\
35 & SST24 J142504.04+345013.7 & 17538304 & 30121 &240,480 & 12.3(11.7) & 165 & PAH,1C\\
36 & SST24 J143316.22+345547.4 &\nodata &\nodata &\nodata &12.3 &215&\nodata\\
37 & SST24 J143734.01+345721.3 &17539840 & 30121 & 240,480& 10.0(9.7) & 19 & featureless\\
38 & SST24 J143641.26+345824.4 &14081280 &20113 & 120,60 &28.1(r) &410 &PAH,1C\\ 
39  & SST24 J143053.70+345836.7 &17537536  & 30121 &120,240 & 81.6(79.7) & 236 &PAH,1B\tablenotemark{f}\\
40 & SST24 J143239.59+350151.5 & 14086656 & 20113 & 120,120 & 10.7 & 197 & PAH,1B\\
41 & SST24 J143544.18+350434.5 & \nodata &\nodata &\nodata &25.8 &95 &Sy 2\tablenotemark{f}\\
42 & SST24 J142614.87+350616.5 &17537792  & 30121 & 240,480& 12.4(11.7) &\nodata  & featureless\tablenotemark{i}\\
43 & SST24 J143518.21+350708.3 & 14081024 &20113 & 120,28 & 233(r) & 3280(r) & PAH,1C\\
44 & SST24 J143046.32+351313.6 & 16347904 & 16 & 240,480 & 16.5(17.5) & 110 & PAH,1B\\
45 & SST24 J143026.13+351923.9 & \nodata &\nodata & \nodata &227(r)\tablenotemark{g} &3290 &AGN\\
46 & SST24 J143039.27+352351.0 & 16349440 & 16 & 120,240 & 33.9(36.9) & 330 & PAH,1C\\
47 & SST24 J143119.79+353418.1 & 14081792 & 20113 & 120,60 & 33.0(r) &415 &PAH,1C\\
48 & SST24 J142714.86+353442.4 &\nodata &\nodata &\nodata &23.4 &\nodata &starburst\tablenotemark{f}\\
49 & SST24 J143121.15+353722.0 &14082048 & 20113 & 120,60 & 49(r) & 900 &PAH,1C\\ 
50 & SST24 J142827.20+354127.7 & 12530432 & 15 & 240,480 & 10.5 & \nodata & Si absorption,2A\tablenotemark{h}\\

\enddata
\tablenotetext{a}{SST24 source name derives from discovery with the
  MIPS 24$\mu$m images; coordinates listed are J2000 24$\mu$m positions with typical 3\,$\sigma$ uncertainty of $\pm$ 1.2\arcsec. AOR number is the observation number in the $Spitzer$ archive.}
\tablenotetext{b}{First value is total integration time in IRS short-low orders 1 and 2; second value is total integration time in IRS long-low orders 1 and 2.}

\tablenotetext{c}{Values of f$_{\nu}$(24$\mu$m) are for an unresolved point source measured from MIPS images, except that sources noted by (r) are resolved within the same galaxy in the 24$\mu$m image, and the flux listed is the total flux for all components. Values in parentheses are the f$_{\nu}$(24$\mu$m) measured independently from the extracted IRS spectra; the mean difference of $\pm$ 4\% between IRS and MIPS values gives an estimate of uncertainties in results.}

\tablenotetext{d}{Values of f$_{\nu}$(70$\mu$m) are for an unresolved point source, measured from MIPS images; fluxes are uncertain for sources 13 and 14 because these are two closely interacting galaxies. Sources without measured f$_{\nu}$(70$\mu$m) are outside of the field of view for the 70$\mu$m survey.}

\tablenotetext{e}{Classification of IRS spectrum, whether showing PAH emission, emission lines, absorption by the 9.7$\mu$m Si feature, emission by the 9.7$\mu$m silicate feature, or no features; numerical value gives classification according to scheme of \citet{spo07}; sources without IRS spectra but with optical classifications are described in footnote f.}

\tablenotetext{f}{SDSS spectra show that source 1 is a type 1 QSO with z = 0.2019, source 3 is a starburst with z = 0.0138, source 8 is a starburst with z = 0.0418, source 10 is a starburst with z = 0.09386, source 11 is a starburst with z = 0.0738, source 13 is a starburst with z = 0.0424, source 14 is a type 1 QSO with z = 1.0820, source 28 is a type 1 QSO with z = 0.7155, source 29 is a starburst with z = 0.0393, source 39 is a Sy 2 AGN with z = 0.084, source 41 is a Sy 2 AGN with z = 0.0287, source 42 is a Sy 1 AGN with z = 0.2168, source 48 is a starburst with z = 0.0773; source 18 is an emission line galaxy within the Kitt Peak International Spectroscopic Survey (KISS) with z = 0.0337. Additional SDSS results for PAH sources that have IRS spectra are included in Table 3.} 

\tablenotetext{g}{spiral galaxy and liner NGC 5656, with SDSS z = 0.01055.}

\tablenotetext{h}{IRS spectrum in \citet{des06}.}

\end{deluxetable}


\clearpage
\pagestyle{empty}

\begin{deluxetable}{lccccccccc} 

\tablecolumns{10}
\tabletypesize{\footnotesize}

\tablewidth{0pc}
\tablecaption{Observed Spectroscopic Properties for New Observations of Sources with PAH Features} 
\tablehead{
  \colhead{Number} & \colhead{Source Name\tablenotemark{a}} &\colhead{f$_{\nu}$(6\ums)\tablenotemark{b}} & \colhead{f$_{\nu}$(15\ums)\tablenotemark{c}}&\colhead{PAH\tablenotemark{d}}  & \colhead {PAH\tablenotemark{d}}& \colhead{[SIV]\tablenotemark{d}} & \colhead{[NeII]\tablenotemark{d}} & \colhead{[NeIII]\tablenotemark{d}} & \colhead{[SIII]\tablenotemark{d}}\\ 
  \colhead{}&  \colhead{} &\colhead{mJy}& \colhead{mJy}& \colhead{6.2\ums }&  \colhead{11.3\ums }& \colhead{10.51\ums}& \colhead{12.81\ums}& \colhead{15.56\ums}& \colhead{33.48\ums}}
\startdata


 2 &1426+3221& 1.31 & 5.82 & 50 & 30 & 1.6 & 23?? &  8.8 & \nodata\\
 4 & 1426+3229 & 1.1 & 4.4 & 42 & 44 & 1.3 & 28 & 1.1 & 17??\\
 6 & 1431+3246 & 1.7 & 6.6 & 68 & 67 & 3.7 & 29 & 12.5 & 19.4\\
 15 & 1426+3333 & \nodata& 8.4 & \nodata & \nodata & \nodata & \nodata &\nodata & \nodata \\
 20 & 1425+3345 & 2.0 & 15.4 & 76 & 76 & $<$4 & 34 & 8.6 & \nodata\\
 23 & 1436+3352 & 1.15 & 5.3 & 80 & 54 & $<$2 & 26 & \nodata & 13.6\\
 25 & 1432+3406\tablenotemark{e} & 1.1 & 2.6 & 15.2 & 15.5 & 2.0 & 3.0 & 6.5 & 2.1\\
 30 & 1431+3438 & 0.7 & 6.8 & 7.7?? & 15.7 & 22.5 & 5.2 & 25.2 & 15.5\\
 31 & 1436+3438 & 9.1 & 35.7 & 30 & 14.3 & $<$1.5 & 17.9 & \nodata \\
 32 & 1431+3445 & 1.8 & 3.8 & 94 & 76 & 1.1 & 33 & \nodata & 5.8\\
 34 & 1425+3446 & 5.0 & 15.5 & 286 & 218 & 9 & 105 & 12.5 & 47\\
 35 & 1425+3450 & 1.92 & 5.65 & 112 & 87 & 1.3 & 37.5 & 4.4 & \nodata\\
 39  & 1430+3458 & 2.40 & \nodata& 36 & 40 & 6.1 & 67 & \nodata & \nodata\\
 44 & 1430+3513 & 1.90 & 7.9 & 63 & 48 & 2 & 12.9 & 4.4 & 7.4 \\
 46 & 1430+3523 & 2.8 & 12.2 & 87 & 71 & $<$1 & 45 & 8.6 & 23 \\

\enddata

\tablenotetext{a}{Source name is truncated name from Table 1.}
\tablenotetext{b}{Continuum flux density at 6\,\um; uncertainties are typically $\pm$ 5\%, based on uncertainties in flux calibration.}
\tablenotetext{c}{Continuum flux density at 15\,\um, uncertainties are typically $\pm$ 5\%, based on uncertainties in flux calibration.}
\tablenotetext{d}{Total flux of feature in units of 10$^{-22}$W cm$^{-2}$, fit with single gaussian; uncertainties deriving from noise in the feature and from flux calibration are typically $\pm$ 10\%, except that features with question marks are weak and are marginal detections; upper limits are 3 $\sigma$.} 
\tablenotetext{e}{IRS spectrum shows [OIV] 25.89\,\um with flux 6.3 in these units.} 

\end{deluxetable}

\clearpage
\pagestyle{empty}

\begin{deluxetable}{ccccccccccccc}
\rotate
\tablecolumns{13}
\tabletypesize{\footnotesize}
\tablewidth{0pt}
\tablecaption{Redshifts, Luminosities and Star Formation Rates of PAH Sources}
\tablehead{
\colhead{No.} & 
\colhead{Source}\tablenotemark{a} &
\colhead{z}\tablenotemark{a} & 
\colhead{z}\tablenotemark{a}& 
\colhead{class}&
\colhead{D$_L$} &
\colhead{6.2\um EW}\tablenotemark{b}&
\colhead{f$_{\nu}$(7.7\ums)\tablenotemark{c}} & 
\colhead{$\nu$L$_{\nu}$\tablenotemark{d}} &
\colhead{log$L_{ir}$} &
\colhead{SFR\tablenotemark{e}} &
\colhead{density\tablenotemark{f}}  & 
\colhead{SFR density\tablenotemark{g}}\\
\colhead{}&
\colhead{} &
\colhead{IRS} &
\colhead{SDSS} &  
\colhead{SDSS}& 
\colhead{Mpc}&
\colhead{\um}& 
\colhead{mJy} &
\colhead{log (ergs s$^{-1}$)} &
\colhead {L$_{\odot}$} &
\colhead{\mdot}&
\colhead{10$^{-5}$Mpc$^{-3}$}&  
\colhead{10$^{-5}$ \mdot Mpc$^{-3}$}
}
\startdata







2 &1426+3221 & 0.2683 & \nodata& \nodata& 1355 &0.65$\pm$0.05 & 9.2(7.9) &44.77   &11.97 &158 & 0.074 & 11.7 \\ 
4 & 1426+3229 & 0.1028 & 0.1008 & starburst & 468 &0.35$\pm$0.03 & 6.9(5.9) & 43.78   &10.98 & 16.2  & 1.09 & 17.7 \\ 
5 & 1426+3244 & 0.1760 & \nodata & \nodata & 842 &0.14$\pm$0.02 & 10.3(8.4) & 44.44&   11.63 & 72 & 0.25&18\\
6 & 1431+3246 & 0.0244 & 0.0227 & starburst & 105 &0.59$\pm$0.04 & 9.8(8.4) &42.67&   9.87&1.25(0.4) &28.5&36\\ 
12 & 1431+3313  & 0.0233 & 0.0226 & starburst & 100 &0.77$\pm$0.03 & 15.1(12.9) &42.81&   10.01&1.7(0.6)  &9.6&16.3\\
15 & 1426+3333 & 0.1514 & \nodata & \nodata & 713 &0.81$\pm$0.02 & 13.0(10.5) & 44.40 & 11.60 & 68 & 0.29 & 19.7\\
16 & 1426+3333 & 0.1548 & \nodata & \nodata & 730 &0.77$\pm$0.03 & 16.4(14.2) & 44.52 & 11.72 & 89 & 0.30 & 26.7\\
19 & 1431+3338 & 0.0329 & \nodata & \nodata & 143 & 0.50$\pm$0.02 & 27.8(24.1) &43.39&   10.59&6.6(0.6) &4.3&28.3\\
20 & 1425+3345 & 0.0715 & 0.0717 &starburst/Sy 2 &319 &0.57$\pm$0.3 & 12.3(8.7) &43.71&   10.91&13.8(3.5) &0.46&6.3 \\ 
23 & 1436+3352 &0.0883 & 0.0866& starburst& 398 & 0.90$\pm$0.05 & 9.4(8.2) &43.78&   10.98&16.2(1.9)  &2.3&37\\
24 &  1425+3402 &0.5642 &\nodata &\nodata & 3270 &0.12$\pm$0.03 & 7.9(6.7) & 45.38&   12.58&646 &0.011&7.1\\
25 & 1432+3406 & 0.0444 & 0.0423 & Sy 2 & 194 & 0.23$\pm$0.04 & 2.2(1.4) &42.54&   9.74&0.93(0.3) &17.9&16.6\\
30 & 1431+3438 &0.0156 & 0.0146 & BCD starburst & 67 & 0.55$\pm$0.09 & 2.7(1.8) &41.72&   8.92&0.14(0.3) &93&13.0\\
31 & 1436+3438 &0.3540 & \nodata& \nodata &1870 & 0.05$\pm$0.01 & 18.9(4.5) &45.33&   12.53&575 &0.0044&2.5\\
32 & 1431+3445 & 0.0835 & 0.0827 &starburst &375 & 0.55$\pm$0.04 & 13.7(12.4) &43.90&   11.1&21.4(2.4) &2.6&55.6\\
34 & 1425+3446 &0.0350 & 0.0344 & starburst & 152 &0.57$\pm$0.02 & 38.6(34.6) &43.58&   10.78&10.2(1.0) &3.3&33.7\\
35 & 1425+3450 & 0.0783 & 0.0768 & starburst & 351& 0.64$\pm$0.02 & 16.6(15.1) &43.92&   11.12&22.4 &2.5&56\\
38 & 1436+3458 & 0.0290 & 0.0302 &starburst& 125 & 0.56$\pm$0.03 &  21.8(18.8) &43.17&   10.37&4.0(0.3)&12.9&51.6\\ 
39  & 1430+3458 &0.0853 & 0.084 & Sy 2 &384 &0.09$\pm$0.01 &  9.2(5.1) &43.74&   10.94&14.8 &0.115&1.7\\
40 & 1432+3501 &0.2371 &0.2357 &starburst & 1180 & 0.37$\pm$0.03 & 9.2(8.0)&44.66& 11.86&123 &0.13&16.0\\
43 & 1435+3507 & 0.0289 & 0.0285 & starburst & 123 &0.66$\pm$0.01  & 89(76)  &43.76 & 10.96 &15.5 & 1.13 & 17.5\\ 
44 & 1430+3513 & 0.0838 & 0.0828 & Sy 2 & 377 & 0.34$\pm$0.03 &  9.7(6.9) &43.75&   10.95&15.1(2.1) &1.34&20.2\\
46 & 1430+3523 &0.0885 & 0.0872 & starburst & 399 &0.57$\pm$0.3 & 14.4(13.0) &43.97&   11.17&25.1(1.5)&0.39&9.8\\
47 &1431+3534 &0.0333&0.0347 &starburst & 144 & 0.72$\pm$0.03 & 39.0(34.6) &43.54&   10.74&9.3(0.8) &7.6&70.7\\
49 & 1431+3537 & 0.0347 & 0.0347 & starburst &150 &0.53$\pm$0.01 & 51.9(48)  &43.70 &  10.9&13.5(0.22) &3.2&43.2\\ 




\enddata

\tablenotetext{a}{Source name is truncated name from Table 1; SDSS redshifts are from optical spectra; IRS redshifts are determined from PAH emission features, assuming rest wavelengths of 6.2$\mu$m, 7.7$\mu$m, 8.6$\mu$m, and 11.3$\mu$m; mean difference between SDSS and IRS redshifts of 0.0012 is estimate of uncertainty in IRS redshift.}
\tablenotetext{b}{Equivalent widths listed (EW) are in the observed frame; equivalent widths in the rest frame are EW/(1+z).} 
\tablenotetext{c}{First number is flux density at peak of 7.7$\mu$m feature; second number is flux density at peak of 7.7$\mu$m feature after subtraction of underlying continuum at 7.7$\mu$m.}
\tablenotetext{d}{Luminosity $\nu$L$_{\nu}$ (7.7$\mu$m) in source rest frame in units of log (ergs s$^{-1}$), determined from peak flux density of 7.7$\mu$m feature without continuum subtraction.}
\tablenotetext{e}{Star formation rate for this source as determined from relation log[SFR] = log[$\nu$L$_{\nu}$ (7.7$\mu$m)] - 42.57; number in parentheses is the SFR that would be derived using the f(H$\alpha$) for KISS sources and the relation log[SFR] = log[L(H$\alpha$)] - 41.1, with no correction for extinction.}
\tablenotetext{f}{Space density for this source determined from inverse of co-moving volume V$_{max}$ which source could occupy and remain within sample f$_{\nu}$(24$\mu$m) $>$ 10 mJy; luminosity distances and co-moving volumes determined by E.L. Wright, http://www.astro.ucla.edu/~wright/CosmoCalc.html, for H$_0$ = 71 \kmsMpc, $\Omega_{M}$=0.27 and $\Omega_{\Lambda}$=0.73. }
\tablenotetext{g}{Star formation rate density produced by this source as determined from $\nu$L$_{\nu}$(7.7$\mu$m).}

\end{deluxetable}

\clearpage
\pagestyle{plaintop}

\begin{deluxetable}{ccccccc} 
\rotate
\tablecolumns{7}
\tabletypesize{\footnotesize}

\tablewidth{0pc}
\tablecaption{Comparison of IRAS and 10 mJy Space Densities and Star Formation Rate Densities}
\tablehead{
 \colhead{log $L_{ir}$} &\colhead{IRAS sources}& \colhead{IRAS space density\tablenotemark{a}} &\colhead{IRAS SFR density\tablenotemark{b}} & \colhead{10 mJy sources\tablenotemark{c}}& \colhead{10 mJy space density\tablenotemark{d}}&\colhead{10 mJy SFR density\tablenotemark{e}}\\ 
\colhead{L$_{\odot}$} &
\colhead{number}&\colhead{10$^{-5}$ Mpc$^{-3}$} &\colhead{10$^{-5}$\mdot Mpc$^{-3}$} &\colhead{number}&\colhead {10$^{-5}$ Mpc$^{-3}$} &  \colhead{10$^{-5}$\mdot Mpc$^{-3}$}
}
\startdata


	









7.5-8.0 & 3 & 11200 & 110 & 0 & \nodata & \nodata\\
8.0-8.5 & 3 & 14000 & 43 & 0 & \nodata & \nodata\\
8.5-9.0 & 9 & 13200 & 130 & 1 & 93 & 13\\
9.0-9.5 & 24 & 670	& 210 & 0 & \nodata & \nodata\\
9.5-10.0 & 69 &329 & 320 & 2 & 46 & 53\\
10.0-10.5 & 168 & 170 & 530 & 2 & 23 & 68\\
10.5-11.0 &157  & 27 & 260 & 10 & 25 & 276\\
11.0-11.5 &122  & 3.9 & 120 & 3 & 5.5 & 121\\
11.5-12.0 &56  & 0.30& 29  & 5(3) & 1.0(0.8) &92(64)\\
12.0-12.5 & 18 & 0.015& 4.5 & 0 & \nodata & \nodata\\
12.5-13.0 &\nodata  & \nodata &\nodata & 2(0) & 0.015 & 9.6(\nodata)\\





\enddata

\tablenotetext{a}{Space densities reproduced from Sanders et al. (2003) for Revised IRAS Bright Galaxy Sample.}
\tablenotetext{b}{SFR density calculated by determining SFR per source from log $L_{ir}$ at center of luminosity interval, related to SFR by log SFR = log $L_{ir}$ - 9.76 (Kennicutt 1998), which is same as relation used to determine SFR per source for 10 mJy sample in Table 3.}
\tablenotetext{c}{Number of sources from Table 3 in each bin of $L_{ir}$; number in parentheses is number that arises after excluding sources with z $>$ 0.2; sources from Table 3 assigned to bins of log $L_{ir}$ by the empirical transformation described in the text, log $L_{ir}$ = log[$\nu$L$_{\nu}$ (7.7$\mu$m)] + 0.78, for $L_{ir}$ in ergs s$^{-1}$.} 
\tablenotetext{d}{Sum of space densities in this bin of log $L_{ir}$ for 10 mJy sample of starbursts in Table 3; sources from Table 3 assigned to bins of log $L_{ir}$ by the empirical transformation described in the text, log $L_{ir}$ = log[$\nu$L$_{\nu}$ (7.7$\mu$m)] + 0.78, for $L_{ir}$ in ergs s$^{-1}$.}
\tablenotetext{e}{Sum of SFR densities in this bin of log $L_{ir}$ for 10 mJy sample of starbursts in Table 3.}

\end{deluxetable}

\clearpage
%
%
\begin{figure}
\figurenum{1}
\includegraphics[scale=.8]{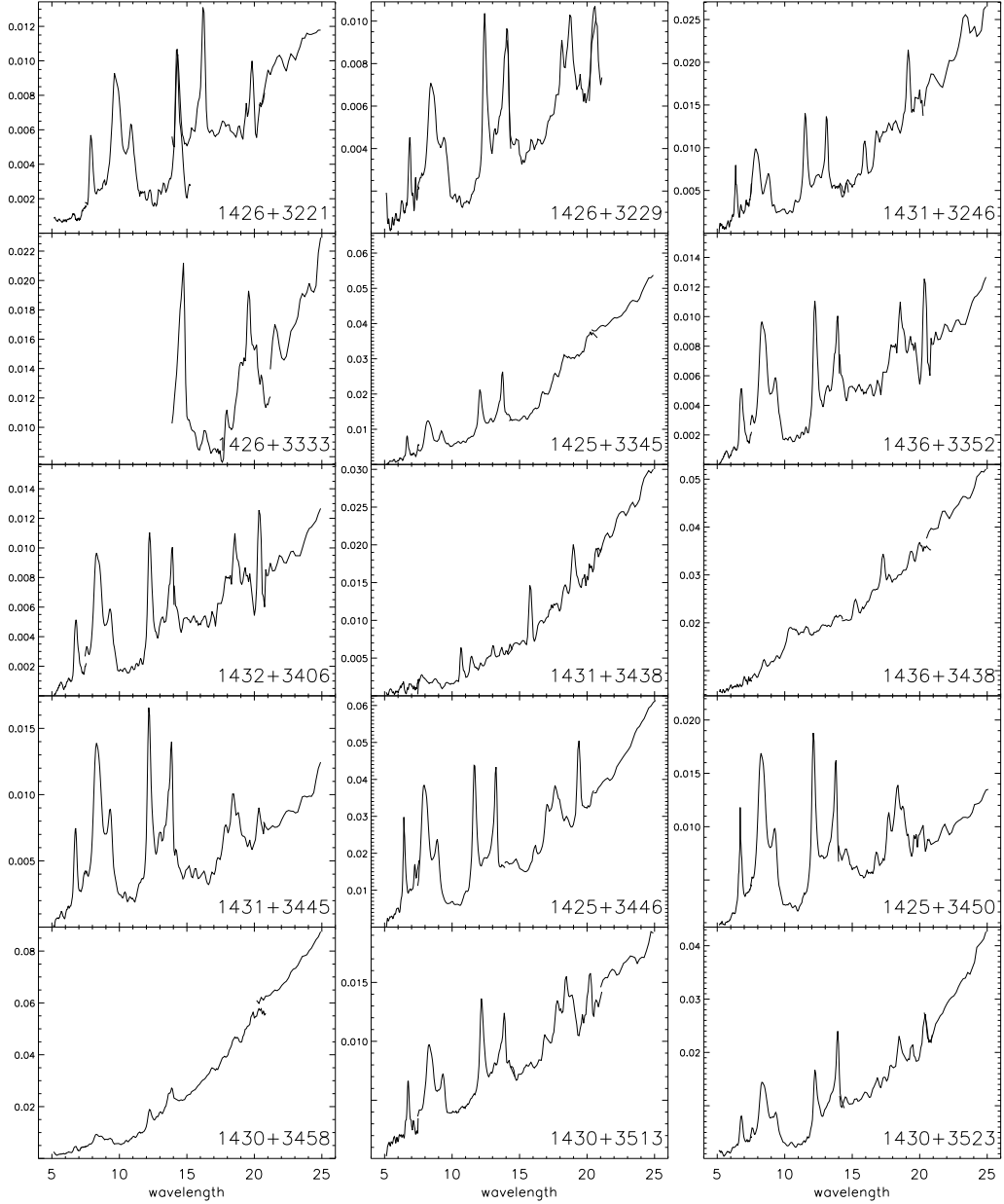}
\caption{Observed spectra for PAH sources in Table 2, identified by truncated source name.  Spectra are not shown beyond 25\,\um because no PAH features are present; flux density units are Jy and wavelength units are \,\um.} 

\end{figure}

\begin{figure}
\figurenum{2}
\includegraphics[scale=1.0]{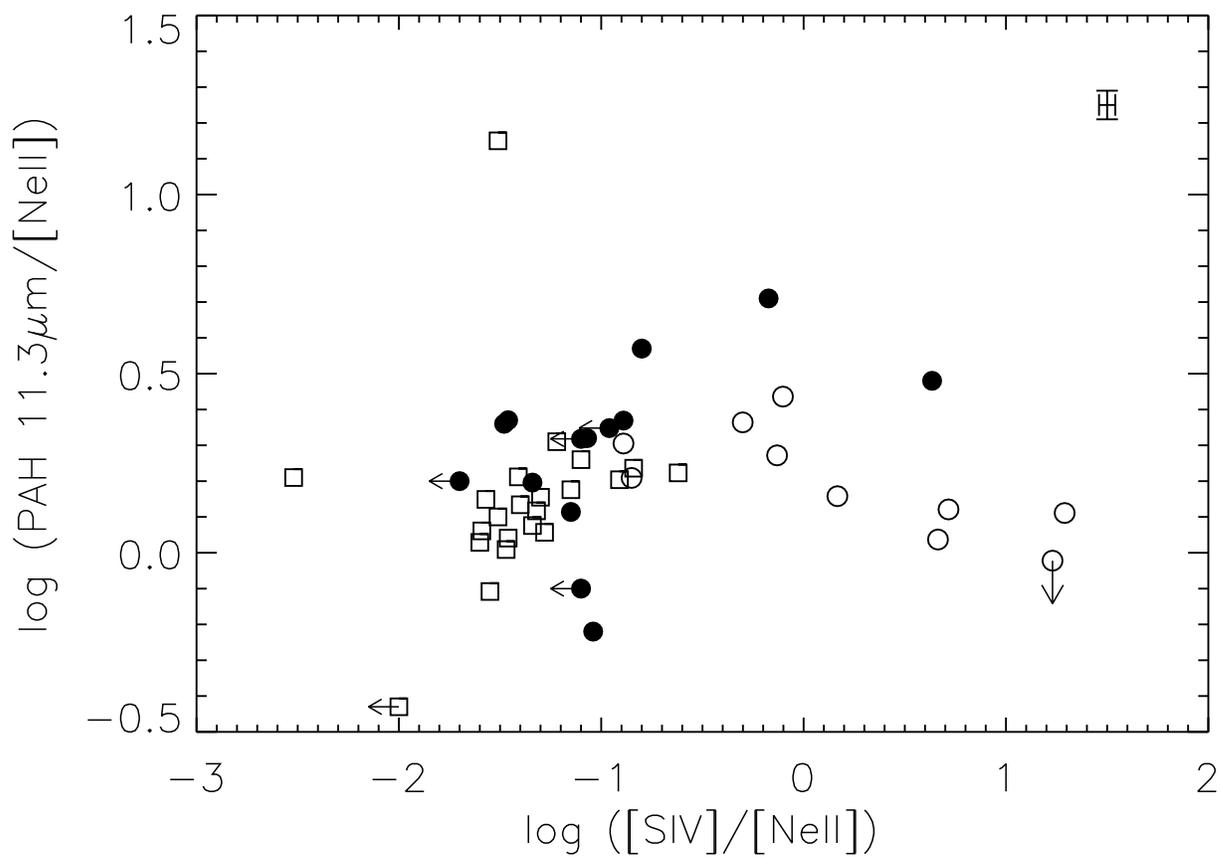}
\caption{Comparison of [NeII] 12.8\,\um and [SIV] 10.5\,\um emission lines with 11.3\,\um PAH strengths; filled circles: PAH sources in current 10 mJy sample; open circles: BCDs from Wu et al. (2006); open squares: starbursts from Brandl et al. (2006). Typical error bars for individual points are shown in upper right of diagram.} 
  
\end{figure}

\begin{figure}
\figurenum{3}
\includegraphics[scale=1.0]{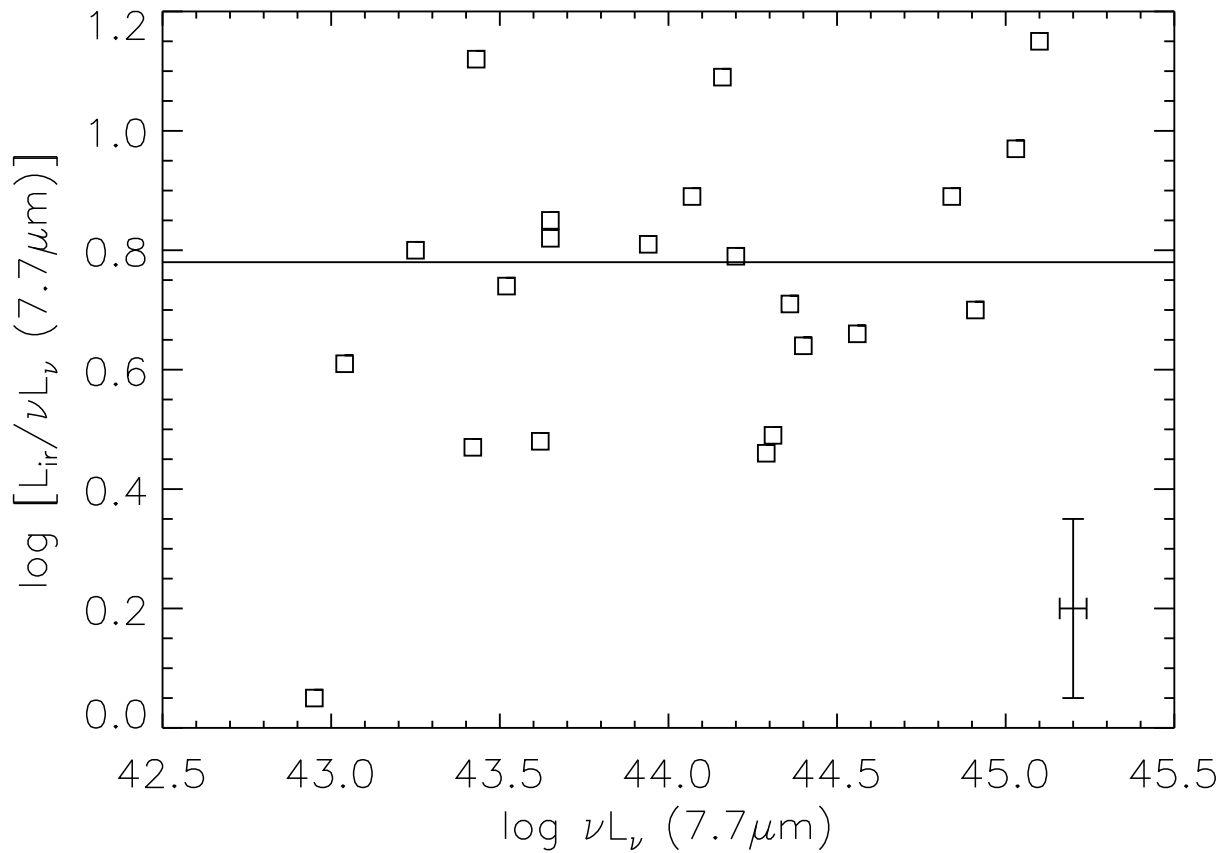}
\caption{PAH luminosity compared to total infrared luminosity $L_{ir}$ from IRAS fluxes for starbursts in \citet{bra06}; fit shown is log $L_{ir}$ = log[$\nu$L$_{\nu}$ (7.7$\mu$m)] + 0.78. The error bar represents the typical factor assumed in Brandl et al. to correct for the fraction of the IRAS flux which is included in the IRS spectrum used to measure f$_{\nu}$ (7.7$\mu$m); uncertainty in this correction factor dominates the measurement uncertainty.}
  
\end{figure}

\newpage
\clearpage

\begin{figure}
\figurenum{4}
\includegraphics[scale=1.0]{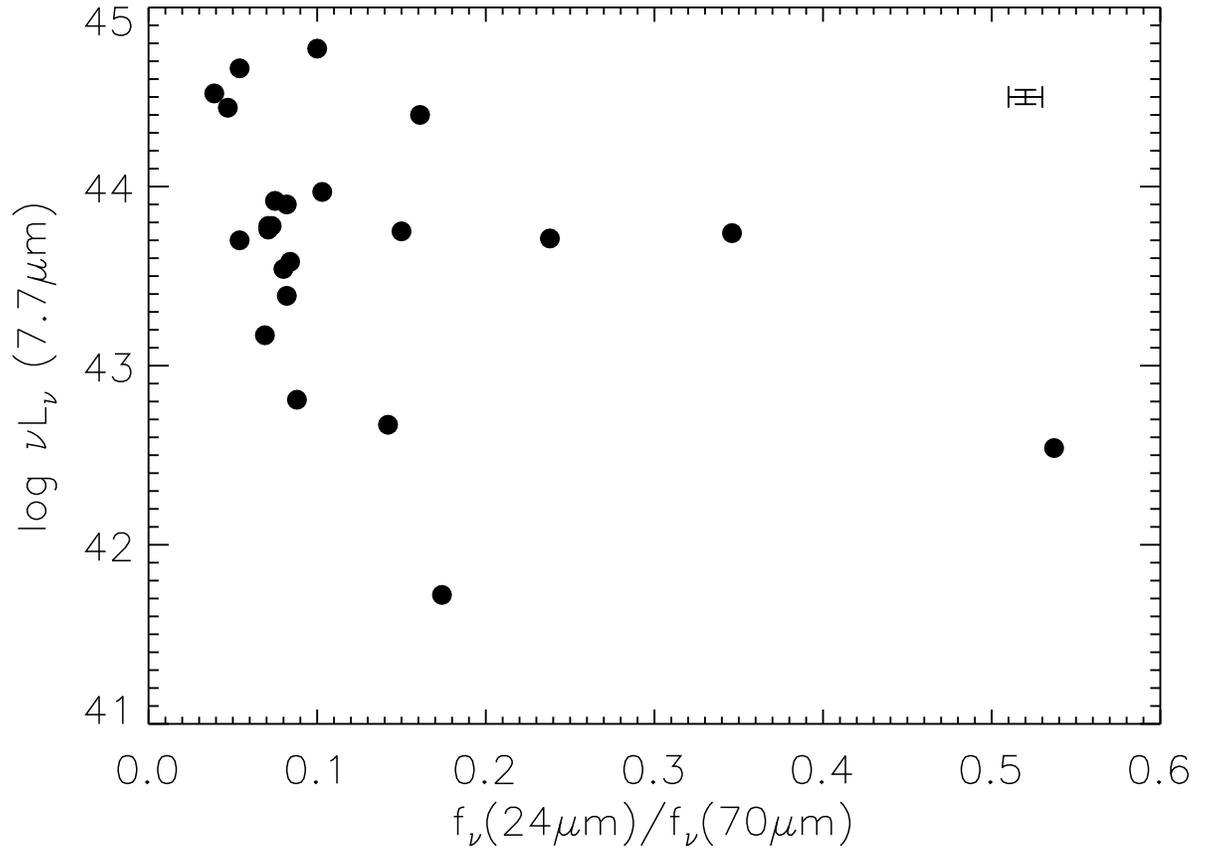}
\caption{Comparison of PAH luminosity and ratio of photometry in 24\,\um and 70\,\um MIPS survey bands for Bootes 10 mJy PAH sources; typical error bars for individual points are shown in upper right of diagram.}
  
\end{figure}
\begin{figure}
\figurenum{5}
\includegraphics[scale=1.0]{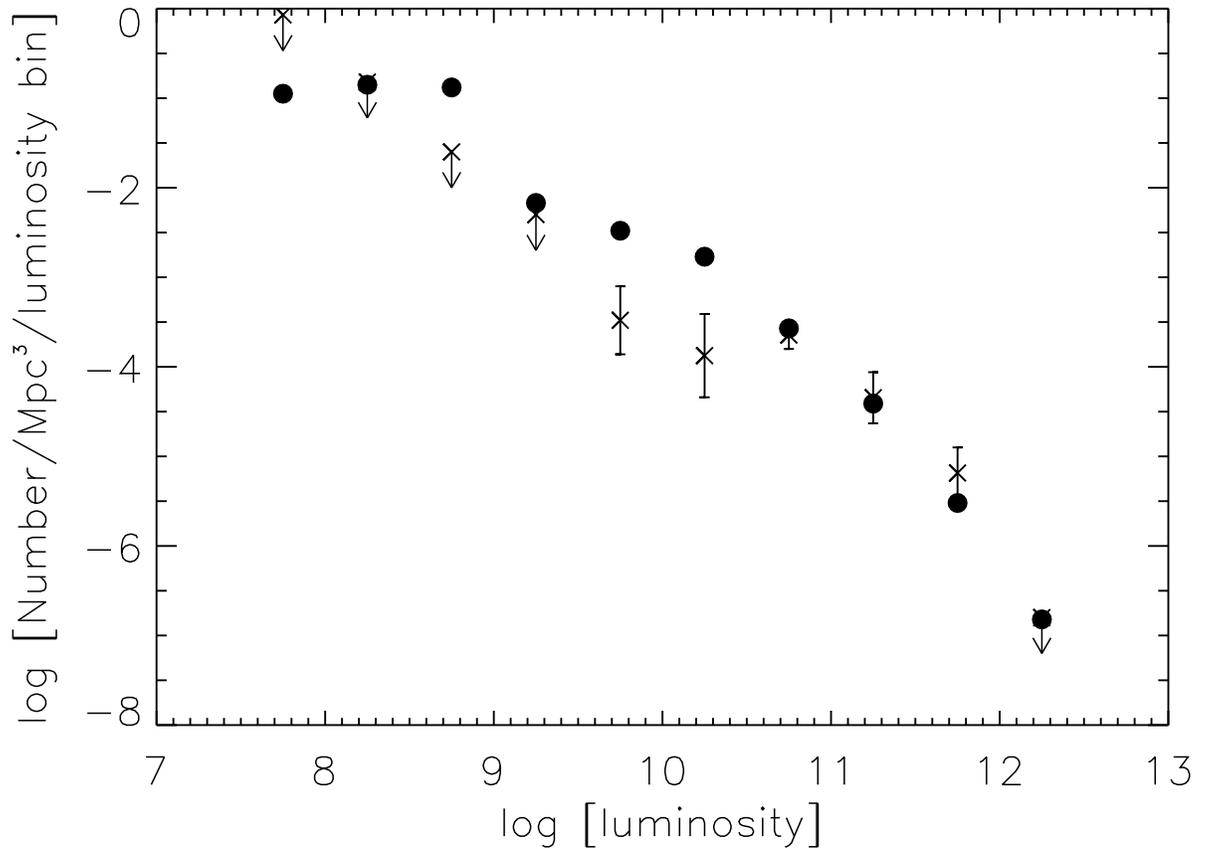}
\caption{Comparison of space densities and luminosities $L_{ir}$ in L$_{\odot}$ from BGS (filled circles) to results from 10 mJy survey in present paper (upper limits and error bars).  Upper limits for bins with no sources assume less than one source detected in that bin; error bars are statistical 1$\sigma$ uncertainties derived from number of sources within each bin in Table 3; statistical error bars for the BGS are smaller than the filled circles.} 
  
\end{figure}

\end{document}